\UseRawInputEncoding
\documentclass[pra,superscriptaddress, notitlepage, reprint,twocolumn]{revtex4-1}
\usepackage[english]{babel}
\usepackage{amsmath,amsthm}
\usepackage{amsfonts}
\usepackage[pdfborder={0 0 0}, colorlinks=true, urlcolor=blue, linkcolor=blue, citecolor=blue]{hyperref}
\usepackage{color}
\usepackage{graphicx}
\usepackage{float}
\usepackage{subfigure}
\usepackage{lipsum}
\usepackage{epstopdf}
\usepackage{dcolumn}
\begin{document}

\title{Quantum thermometry in diffraction-limited systems}
\author{Dong  Xie}\email{xiedong@mail.ustc.edu.cn}

\affiliation{College of Science, Guilin University of Aerospace Technology, Guilin, Guangxi 541004, People's Republic of China}
\affiliation{State Key Laboratory for Mesoscopic Physics, School of Physics, Frontiers Science Center for Nano-Optoelectronics, and
Collaborative Innovation Center of Quantum Matter, Peking University, Beijing 100871, People's Republic of China}
\author{Chunling Xu}
\affiliation{College of Science, Guilin University of Aerospace Technology, Guilin, Guangxi 541004, People's Republic of China}
\author{An Min Wang}
\affiliation{Department of Modern Physics, University of Science and Technology of China, Hefei, Anhui 230026, People's Republic of China}

\begin{abstract}
  We investigate the ultimate quantum limit of resolving the temperatures of two thermal sources affected by the diffraction. More quantum Fisher information can be obtained with the priori information than that without the priori information.  We carefully consider two strategies: the simultaneous estimation and the individual estimation.  The simultaneous estimation of two temperatures is proved to satisfy the saturation condition of quantum Cram\'{e}r bound and performs better than the individual estimation in the case of small degree of diffraction given the same resources. However, in the case of high degree of diffraction, the individual estimation performs better. In particular, at the maximum diffraction, the simultaneous estimation can not get any information, which is supported by a practical measurement, while the individual estimation can still get the information. In addition, we find that for the individual estimation, a practical and feasible estimation strategy by using the full Hermite-Gauss basis can saturate the quantum Cram\'{e}r bound without being affected by the attenuation factor at the maximum diffraction.
\end{abstract}
\maketitle

\section{Introduction}
In classical optics, optical imaging resolution is limited by the diffraction. For over a century, Rayleigh's criterion had been used as a limit of  resolution of two incoherent point sources\cite{lab1,lab2}. In  the last decade, the limit can be beaten by a
variety of superresolution techniques, such as, fluorescence microscopy\cite{lab3,lab4,lab5}.

Tang \textit{et.al.}\cite{lab6} first investigated the imaging resolution limit with the tool in quantum metrology. They obtained the lower bound of the separation between two incoherent point sources and  showed  that the spatial-mode demultiplexing can approach the optimal measurement, which is superior to direct measurement. This seminal work opened up a wide range of interest in exploring quantum imaging using quantum Fisher information (QFI).  They mainly extended the superresolution technique to deal with two-dimensional\cite{lab7} and three-dimensional imaging\cite{lab8,lab9,lab10,lab11}, many sources\cite{lab12,lab13,lab14,lab15,lab16}, the effects of noise\cite{lab17,lab18}, and  the optimal measurement for the practical superresolution imaging\cite{lab19}.

Up to now, very little work has been done to investigate the effect of diffraction on quantum thermometry, which mainly involves improving precision standards for temperature sensing in the quantum regime\cite{lab20}.  Improving temperature measurement precision is important in the quantum thermodynamics and modern quantum technology\cite{lab21,lab22,lab23}.
 The commercially available pyrometer is one of the most common noncontact thermometer, which is the measurement of the thermal infrared radiation naturally emitted by all heated samples\cite{lab24,lab25}. Like the quantum imaging, it is necessary to study the effect of diffraction on temperature measurement precision for obtaining the optimal temperature measurement.

In this article, we fill in the gaps above. We investigate the ultimate quantum limit of resolving the temperatures of two thermal sources
affected by the diffraction.
When one knows a priori that the two temperatures are always the same, the maximum diffraction reduces the QFI of the high temperature by
half and the diffraction has little effect on the measurement of the low temperature. We find that the prior information can help to obtain twice as much QFI as without the priori information (the two temperatures are independent). More importantly, we find that the simultaneous estimation is superior to the individual estimation in the case of small degree of diffraction. In the case of high degree of diffraction, the individual estimation can perform better. In addition, we utilize a practical and feasible estimation strategy based on the optimized error transfer formula to obtain the individual temperature estimation uncertainty, which can saturate the quantum Cram\'{e}r bound (QCRB) at the maximum diffraction. Finally, we show that the diffraction will reduce the precision of
the simultaneous estimation with a practical measurement operator, which can not obtain any information
at the maximum diffraction.

This article is organized as follows. In Section II, we introduce the imaging model and the density matrix in which temperature information is encoded. In Section III, we obtain the QFI when the two thermal sources have the same temperature. In Section IV, the simultaneous estimation and the individual estimation are used to obtain the QFI, and compare the merits of the two strategies. In Section V, we investigate a practical and feasible estimation of the single parameter. The simultaneous estimation with a practical measurement operator is studied in  Section VI. We make a brief conclusion in Section VII.

\section{the imaging model}
\begin{figure}
   \includegraphics[scale=0.3]{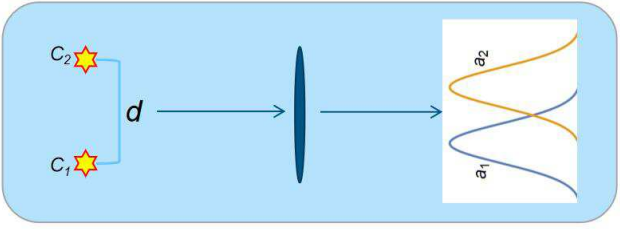}
  \caption{\label{fig.1}Schematic diagram of the diffraction-limited imaging of thermal states from two pointlike sources. Two
  source's modes $c_{1(2)}$ describe the light from the two thermal sources with the distance $d$, which are populated with photon numbers $N(1\pm \gamma)$. $a_{1(2)}$ are the nonorthogonal imaging modes due to the diffraction.}
\end{figure}

We consider the model of a linear optical imaging system in the far field, as shown in Fig.~\ref{fig.1}.
Two thermal pointlike sources are monochromatic with the frequency $\omega$ and located in the object plane, orthogonal
to the optical axis, at position $-d/2$ and $d/2$.  We define that $T_1$ and  $T_2$ are temperatures of the two sources associated with the
field operators $c_1$ and $c_2$, respectively.
We assume that the two sources emit a total mean photon number equal to $2N$, where $N=1/2[1/(\chi_1-1)+1/(\chi_2-1)]$ with $\chi_i=e^{\omega/T_i}$(the reduced Planck constant $\hbar=1$ and Boltzmann constant $\kappa_B=1$ throughout this article). The sources can be described by the density matrix $\rho_0=\rho_{c_1}[(1-\gamma)N]\otimes\rho_{c_2}[(1+\gamma)N]$, where $\gamma=(\chi_1-\chi_2)/(\chi_1+\chi_2)$ takes into account the possibly different temperatures of the two sources.
In the Glauber-Sudarshan P-representation, the density matrix can be also described by
   \begin{align}
\rho_0=\int d^2\alpha_1d^2\alpha_2 P_{c_1,c_2}(\alpha_1,\alpha_2)|\alpha_1,\alpha_2\rangle\langle\alpha_1,\alpha_2|,\label{eq:A01}
\end{align}
where $|\alpha_{1}\rangle$ and $|\alpha_{2}\rangle$ are coherent states of the field operators $c_{1}$ and $c_{2}$ respectively, and the Glauber-Sudarshan
$P$ function is $P_{c_1,c_2}(\alpha_1,\alpha_2)=P_{c_1}(\alpha_1)P_{c_2}(\alpha_2)$, with
\begin{align}
 P_{c_1,c_2}(\alpha_1,\alpha_2)=\frac{1}{\pi^2N^2(1-\gamma^2)}e^{[-|\alpha_1|^2/(1-\gamma)-|\alpha_2|^2/(1+\gamma)]}.
\end{align}

  The point-spread function $\psi(x)$ determines the field operators on the image plane, which read
\begin{align}
a_1^\dagger=\int dx \psi(x+d/2)a_x^\dagger,\ \
a_2^\dagger=\int dx \psi(x-d/2)a_x^\dagger,
\end{align}
where $a_x^\dagger$ is the canonical creation operator for a field localized at position $x$ on the image screen.

A diffraction-limited optical system transforms the source operators as\cite{lab26}
\begin{align}
c_1\longrightarrow\sqrt{\eta}a_1+\sqrt{1-\eta}v_1,\label{eq:A04}\\
c_2\longrightarrow\sqrt{\eta}a_2+\sqrt{1-\eta}v_2,
\label{eq:A05}
\end{align}
where $\eta$ is an attenuation factor, $v_1$ and $v_2$ are auxiliary environmental modes in the vacuum state.

The operators $c_1^\dagger$ and $c_2$ do not commute due to the nonzero overlap between the two point-spread functions $\psi(x+d/2)$ and $\psi(x-d/2)$.  To obviate this problem, the orthonormal image modes are introduced
\begin{align}
\psi_\pm(x)=\frac{\psi(x+d/2)\pm\psi(x-d/2)}{\sqrt{2(1\pm s)}},
\end{align}
where $s$ is the overlap between the source images
\begin{align}
s=\int d^2x\psi^*(x+d/2)\psi(x-d/2).
\end{align}
$s$ quantifies the diffraction introduced by the imaging optical system. $s=1$ represents the maximum diffraction. $s=0$ means that there's no diffraction.
By taking the sum and difference of the relations in Eq.~(\ref{eq:A04}) and Eq.~(\ref{eq:A05}), one can obtain
\begin{align}
c_\pm:=\frac{c_1\pm c_2}{\sqrt{2}}\rightarrow\sqrt{\eta_\pm}a_\pm+\sqrt{1-\eta_\pm}v_\pm,
\label{eq:A08}
\end{align}
where $a_\pm=(a_1\pm a_2)/\sqrt{2(1\pm s)}$ are orthogonal symmetric and antisymmetric mode operators associated with the modes $\psi_\pm(x)$, the effective attenuation factors are $\eta_\pm=\eta(1\pm s)$ and $v_\pm$ associated with auxiliary modes in the vacuum state.
Inverting Eq.~(\ref{eq:A08}), we can write
\begin{align}
a_\pm:=\sqrt{\eta_\pm}c_\pm+\sqrt{1-\eta_\pm}v_\pm.
\label{eq:A09}
\end{align}
The density matrix in the image plane can be obtained by using Eq.~(\ref{eq:A09}) to propagate the quantum state of source in Eq.~(\ref{eq:A01})\cite{lab27}, as shown in Appendix,
\begin{align}
\rho=\int d^2\alpha_+d^2\alpha_- P_{a_+,a_-}(\alpha_+,\alpha_-)|\alpha_+,\alpha_-\rangle\langle\alpha_+,\alpha_-|,\label{eq:A10}
\end{align}
where the corresponding $P$ function is
\begin{align}
 P_{a_+,a_-}(\alpha_+,\alpha_-)=\frac{1}{\pi^2\textmd{det} V}e^{-{\mathbf{A}}^\dagger V^{-1}\mathbf{A}},
\end{align}
with the definition
$\mathbf{A}=(\alpha_+,\alpha_-)^T$ and
 \[
V= \left(
\begin{array}{ll}
N_+\ \ \ \ \ \ \ \ \ \ \ \gamma\sqrt{N_+N_-}\\
\gamma\sqrt{N_+N_-}\ \ \ N_-\\
  \end{array}
\right ),
  \]
in which,
$N_\pm=N \eta(1\pm s)$.
\section{Two thermal sources  with the same temperature}
We first consider that temperatures of the two sources are always the same, i.e., $T_1=T_2=T$. According to Eq.(11), The density matrix in the image plane is a product state, which can be described in the number-diagonal states of the form
\begin{align}
\rho=\rho_+\otimes\rho_-,
\end{align}
with the density matrixs associated with the field operators $a_\pm$
\begin{align}
\rho_\pm=\sum_{n=0}^\infty p_\pm(n)|n\rangle_\pm\langle n|
\end{align}
where
\begin{align}
 p_\pm(n)=\frac{(M_\pm)^n}{(M_\pm+1)^{n+1}},
\end{align}
$M_\pm=\frac{\eta(1\pm s)}{e^{\omega/T}-1}$, and $|n\rangle_\pm=\frac{1}{\sqrt{n!}}{(a^\dagger_\pm)^n}|0\rangle$ denote Fock states with $n$ photons in the image plane.
Due to that it is diagonal state, the QFI of the temperature $T$ can be directly calculated
\begin{align}
& \mathcal{F}(T)=\sum_{n=0}^\infty \frac{[\partial_Tp_+(n)]^2}{p_+(n)}+\frac{[\partial_Tp_-(n)]^2}{p_-(n)}\\
 &=\frac{2\chi^{2}\omega^2\eta(\chi-1+\eta-s^2\eta)}{(\chi-1)^2T^4(-1+\chi+\eta-s \eta)(-1+\chi+\eta+s \eta)},
\end{align}
where the short hand $\partial_T=\frac{\partial}{\partial T}$ and $\chi=e^{\omega/T}$.

At low temperature $\omega/T\gg1$, we can achieve

\begin{align}
 \mathcal{F}(T)=\frac{2\omega^2\eta}{T^4e^{\omega/T}}.
\end{align}
It is independent of the degree of diffraction $s$, which can show that the diffraction has little effect on the measurement of low temperature.

At high temperature $\omega/T\ll1$, we can obtain
\begin{align}
 \mathcal{F}(T)\approx\frac{2\eta[\omega/T+\eta(1-s^2)]}{T^2[\omega/T+\eta(1-s)][\omega/T+\eta(1+s)]}
\end{align}
In this case, we find that $\mathcal{F}(T)|_{s=1}/\mathcal{F}(T)|_{s=0}=1/2$. It means that the maximum diffraction reduces the QFI by half. \begin{figure}
   \includegraphics[scale=0.5]{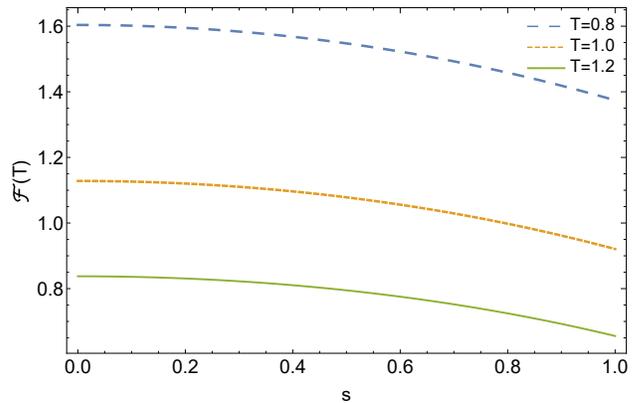}
   \caption{\label{fig.2}This plot shows the QFI $\mathcal{F}(T)$, computed from Eq.16, versus the degree of diffaration $s$. The dimensionless parameters are given by $\omega=1$, $\eta=0.5$. }
\end{figure}

In the general case, we can see that the diffraction will reduce the QFI of temperature as shown in Fig.~\ref{fig.2}.
At the maximum diffraction, we still obtain a finite QFI. It demonstrates that the diffraction has no great influence on temperature measurement in the case of the two thermal sources with the same temperature.
\section{estimating two different temperatures}
In this section, we want to estimate the temperatures $T_1$ and $T_2$ of the two thermal sources. In this case, the two temperatures are independent.
The estimation precision of $(T_1, T_2)$, governed by its covariance matrix $\textmd{Cov}(T_1, T_2)$, is lower bounded via QCRB\cite{lab278}
\begin{align}
\textmd{ Cov}(T_1, T_2)\geq(\nu \mathcal{H})^{-1}
\end{align}
where $\mathcal{H}$ is the QFI matrix and $\nu$ denotes the classical contribution from repeating the
experiment.
There are two measurement strategies: one is the simultaneous estimation of the two temperatures, the other is the individual estimation of the two temperatures. A lot of works\cite{lab28,lab29,lab30,lab31,lab32,lab33,lab34,lab35,lab36,lab37} clearly showed that the simultaneous estimation can be more precise than the individual estimation given by the same resource. And then we're going to look at whether that's true in the diffraction case.

For the simultaneous estimation, the total estimation uncertainty of the two temperatures is given by
\begin{align}
 (\delta^2T_1+\delta^2T_2)|_{\textmd{sim}}=\textmd{tr}[\textmd{Cov}(T_1, T_2)]\geq\textmd{tr}(\nu \mathcal{H})^{-1}\nonumber\\
 =\frac{1}{\nu}\frac{\mathcal{H}^{11}+\mathcal{H}^{22}}{\mathcal{H}^{11}\mathcal{H}^{22}-|\mathcal{H}^{12}|^2}
 \label{eq:A20}
\end{align}
where $\mathcal{H}^{ij}$ ($i,j=1,2$) represent the elements of the QFI matrix $\mathcal{H}$.

For the individual estimation, the estimation uncertainties of the the two temperatures are given by
\begin{align}
\delta^2T_1|_{ind}\geq\frac{1}{\nu/2}\frac{1}{\mathcal{H}^{11}},\label{eq:A21}\\
\delta^2T_2|_{ind}\geq\frac{1}{\nu/2}\frac{1}{\mathcal{H}^{22}},\label{eq:A22}
\end{align}
where we consider $T_1$ and $T_2$ are individually measured $\nu/2$ times so that the total number of measurements is consistent with the case of the simultaneous estimation.
In the case of the individual estimation, the lower bound in Eq.~(\ref{eq:A21}-\ref{eq:A22}) can be saturated with the large number of repeated measurements ($\nu\gg1$).

In the case of the simultaneous estimation, the lower bound in Eq.~(\ref{eq:A20}) is saturated by satisfying the weak commutation relation in addition to the large number of repeated measurements, which is described as\cite{lab38}
\begin{align}
\textmd{tr}[\rho[\mathcal{L}_1,\mathcal{L}_2]]=0,
\end{align}
where $\mathcal{L}_i$ ($i=1,2$) are the symmetric logarithmic derivatives, which are defined as operator solutions to equations $\partial_i\rho=\frac{1}{2}(\mathcal{L}_i\rho+\rho\mathcal{L}_i)$, where $\partial_i = \partial_{T_i}$ denotes partial derivative with respect to the $i$¡¯th element of the vector of estimated parameters $(T_1,T_2)$.

The quantum state $\rho$ is a Gaussian state.
For the Gaussian state,
the QFI matrix $\mathcal{H}$ and symmetric logarithmic derivatives can be described as\cite{lab39}
\begin{align}
\mathcal{H}^{ij}=1/2\textmd{vec}[\partial_i\sigma]^\dagger \mathcal{R}^{-1}\textmd{vec}[\partial_j\sigma]+2\partial_i\mathbf{d}^\dagger\sigma^{-1}\partial_j\mathbf{d},\\
\mathcal{L}_i=\Delta \mathbf{A}^\dagger \mathcal{R}^{-1}\textmd{vec}[\partial_i\sigma]\Delta \mathbf{A}-\frac{1}{2}\textmd{tr}[\sigma\mathcal{R}^{-1}\textmd{vec}[\partial_i\sigma]]\nonumber\\
+2\Delta \mathbf{A}^\dagger\sigma^{-1}\partial_i\mathbf{d},
\end{align}
where the elements
of the displacement vector $\mathbf{d}$ and the covariant matrix $\sigma$ are defined as $d_i=\textmd{tr}[\rho A_i]$ and $\sigma_{ij}=\textmd{tr}\{\rho{\Delta A_i,\Delta A_j}\}$, $\mathbf{A}=(a_+,a_-,a_+^\dagger,a_-^\dagger)^T$, $\Delta A_i=A_i-d_i$, $\mathcal{R}^{-1}=\bar{\sigma}\otimes\sigma-\mathbf{K}\otimes\mathbf{K}$, and $\mathbf{K}=\textmd{diag}(1,1,-1,-1)$. Bar as in $\sigma$ denotes the complex conjugate, $\{.,.\}$ denotes the anticommutator.   vec[.]
denotes vectorization of a matrix, which is defined as a column vector constructed from columns of a matrix. By calculation, the variance matrix can be achieved\\
$\sigma=$
 \[
 \left(
\begin{array}{ll}
2N_++1\ \ \ \ \ \ \ -2\gamma\sqrt{N_+N_-}\ \ \ \ \ \ 0\ \ \ \ \ \ \ \ \ \ \ \ \ 0\\
-2\gamma\sqrt{N_+N_-}\  \ \ \ \ \ 2N_-+1\ \ \ \ \ \ \ \ 0\ \ \ \ \ \ \ \ \ \ \ \  \ 0\\
\ \  \ \ \ \ 0\ \ \ \ \ \ \ \ \ \ \ \ \ \ \ \ \ \ \  \ \ 0\ \ \ \ \ \ 2N_++1\ \ \ -2\gamma\sqrt{N_+N_-}\\
\  \ \ \ \ \ 0\ \ \ \ \ \ \ \ \ \ \ \ \ \ \ \  \ \ \ \ \ 0\ \ -2\gamma\sqrt{N_+N_-}\ \ \ \ \ \ 2N_-+1
  \end{array}
\right ).
  \]

\subsection{Increasing QFI with the priori information}

 When we know a priori that the two temperatures of the two sources are always equal, i.e., $T_1=T_2$. With the priori information, the saturated uncertainty of $T_1$  is given by
 \begin{align}
\delta^2T_1|_{\textmd{pri}}=\frac{1}{\nu/2}\frac{1}{\mathcal{F}(T_1)},
\end{align}
 where the QFI $\mathcal{F}(T_1)$ is described in Eq.16.

Without the priori information, when $T_2\rightarrow T_1$, we can obtain the analytical results of the QFI matrix based on Eq.(24)
\begin{align}
&\mathcal{H}^{11}(T_2\rightarrow T_1)=\mathcal{H}^{22}(T_2\rightarrow T_1)=\nonumber\\
&\frac{\chi_1^2\omega^2\eta}{T_1^4(\chi_1-1)^2[(1-\chi_1-\eta)^2-s^2\eta^2](-1+\chi_1+\eta-s^2\eta)}\times\nonumber\\
&[(1+\chi_1^2)(s^2-2)-2(s^2-1)^2\eta^2-4(s^2-1)\eta+\nonumber\\
&\chi_1(4-4\eta+s^2(4\eta-2))].
\end{align}
In this case, the uncertainty of the temperature $T_1$ is
\begin{align}
\delta^2T_1|_{ind}=\frac{1}{\nu/2}\frac{1}{\mathcal{H}^{11}(T_2\rightarrow T_1)},
\end{align}
\begin{figure}
  \includegraphics[scale=0.45]{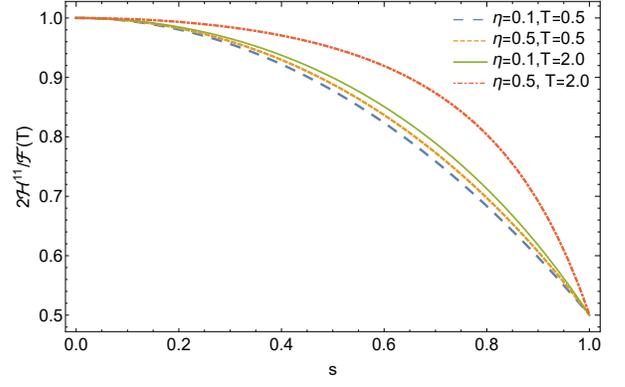}
  \caption{\label{fig.3}This plot shows the QFI $\mathcal{F}(T)$, computed from Eq.16, versus the degree of diffaration $s$. The dimensionless parameters are given by $\omega=1$, $\eta=0.5$. }
\end{figure}

When there is no diffraction ($s=0$), $\mathcal{F}(T_1)=\mathcal{H}^{11}(T_2\rightarrow T_1)+\mathcal{H}^{22}(T_2\rightarrow T_1)=2\mathcal{H}^{11}(T_2\rightarrow T_1)$. However, when there is diffraction ($s\neq0$), $\mathcal{F}(T_1)>2\mathcal{H}^{11}(T_2\rightarrow T_1)$, as shown in Fig.~\ref{fig.3}. In particular, when $s=1$, $\mathcal{F}(T_1)=2(\mathcal{H}^{11}+\mathcal{H}^{22})=4\mathcal{H}^{11}$. It shows that more QFI can be obtained with the priori information of $T_1=T_2$ than that without the priori information when subjected to diffraction. At the maximum diffraction ($s=1$), the prior information can help to obtain twice as much QFI as without the priori information.

\subsection{Simultaneous estimation versus individual estimation}
\begin{figure}
   \includegraphics[scale=0.25]{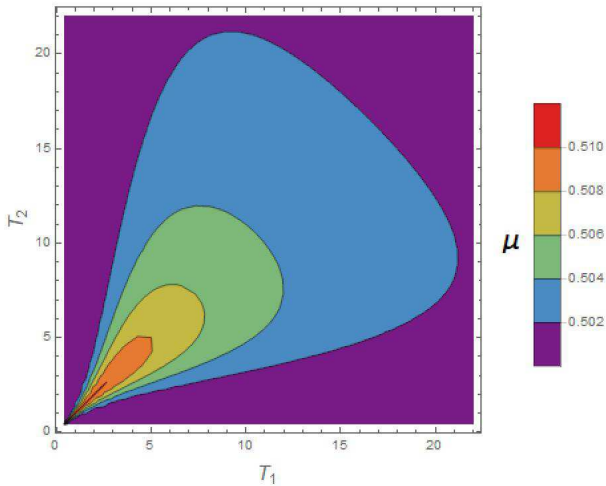}
  \caption{\label{fig.4}Contourplot shows the ratio of the simultaneous uncertainty
   to the individual uncertainty $\mu$ versus the two temperatures $T_1$ and $T_2$. The dimensionless parameters are given by $\omega=10$, $\eta=0.5$, and $s=0.5$. }
\end{figure}

\begin{figure}
   \includegraphics[scale=0.52]{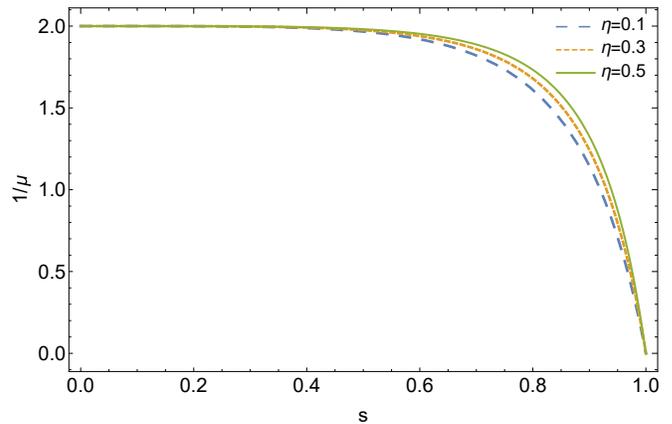}
  \caption{\label{fig.5}Diagram shows the ratio between the individual uncertainty
  and the simultaneous uncertainty, $1/\mu$ versus the degree of diffraction $s$. The dimensionless parameters are given by $\omega=10$, $T_1=8$, and $T_2=10$. }
\end{figure}

\begin{figure}
  \includegraphics[scale=0.6]{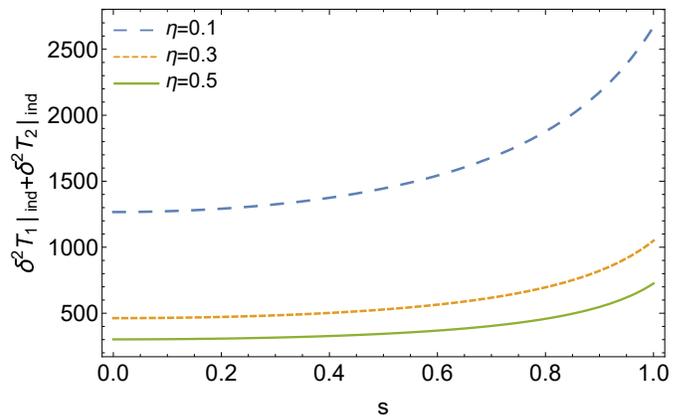}
  \caption{\label{fig.6}Diagram shows the individual uncertainty
  $\delta^2T_1|_{ind}+\delta^2T_2|_{ind}$ versus the degree of diffraction $s$. Here, the values of the  selected parameters are the same as in Fig.~\ref{fig.5}.}
\end{figure}

For simultaneous estimation, we show that the lower bound in Eq.(20) can be saturated by analytically deriving
\begin{align}
\textmd{tr}[\rho[\mathcal{L}_1,\mathcal{L}_2]]=
\textmd{vec}[\partial_1\sigma]^\dagger\mathcal{R}^{-1}(\bar{\sigma}\otimes\mathbf{K}-\mathbf{K}\otimes\sigma)\mathcal{R}^{-1}\textmd{vec}[\partial_2\sigma]\nonumber\\
+4\partial_1\mathbf{d}^\dagger\sigma^{-1}\mathbf{K}\sigma^{-1}\partial_2\mathbf{d}=0.
\end{align}
From now on, we set $\nu = 1$ for the sake of convenience due to that this article is independent of the number of measurements.
The QFI matrix can be analytically derived by Eq.(24). However, the general form is verbose. Results are presented by using numerical values, as shown in Fig.~\ref{fig.4}-\ref{fig.6}.
We define the factor $\mu$ as the ratio of the simultaneous uncertainty and the individual uncertainty, i.e.,
 \begin{align}
\mu=\frac{ (\delta^2T_1+\delta^2T_2)|_{\textmd{sim}}}{\delta^2T_1|_{ind}+\delta^2T_2|_{ind}}
=\frac{\mathcal{H}^{11}\mathcal{H}^{22}}{\mathcal{H}^{11}\mathcal{H}^{22}-|\mathcal{H}^{12}|^2},
\end{align}
where the latter equation comes from the saturated QCRB.

From Fig.~\ref{fig.4}, we can see that in the case of $s=0.5$, the simultaneous estimation uncertainty  is less than the individual uncertainty given by the same resource, i.e., the ratio factor $\mu<1$. It shows that the simultaneous estimation performs better than the individual estimation. When the temperature difference ($|T_1-T_2|$) is relatively large or both temperatures are relatively high ($T_1\gg\omega$ and $T_2\gg\omega$), we find that the ratio $\mu$ is close to 1/2. It indicates that simultaneous estimation in this case is a better use of resources to improve measurement precision.

From Fig.~\ref{fig.5}, we can see that the ratio of the individual uncertainty and the simultaneous uncertainty, $1/\mu$, decreases with the increase of $s$. In particular, the ratio $1/\mu$ approaches 0 as the diffraction degree approaches 1. It indicates that the advantage of the simultaneous estimation decreases as $s$ increases. At the maximum diffraction, the simultaneous estimation uncertainty will be infinite, which means that the maximum diffraction completely prevents the simultaneous estimation from obtaining the information of both temperatures.  In addition, we can see that the  attenuation factor $\eta$ has very little effect on the ratio, especially if $s$ is around 0 and 1.

As shown in Fig.~\ref{fig.6}, although the individual estimation uncertainty ($\delta^2T_1|_{ind}+\delta^2T_2|_{ind}$) also increases with $s$, it is always finite. It means that the individual estimation can obtain the information of the two temperatures when subjected to the maximum diffraction.

\section{a practical and feasible estimation of the single parameter}
A simple way to measure the individual estimation error of the single parameter $T_i $ is given by the error transfer formula\cite{lab40,lab41}

\begin{align}
(\delta T_i)^2=(\delta {X})^2/(\partial_i\langle{X}\rangle)^2,
\end{align}
where $(\delta {X})^2=\langle X^2\rangle-\langle X\rangle^2$, and $\langle \bullet\rangle=\textmd{tr}[\bullet \rho]$. It just needs to measure the average value of a single measurement observable $X$.

For a single parameter, Gessner \textit{et.al.}\cite{lab42} provided an analytical optimization over all possible linear combinations of some given possible measurement observables $\mathbf{X}=(X_1,...,X_K)^T$.

With the optimal  linear combinations ${X}_{\textbf{m}}={\textbf{m}}\cdot\textbf{X}\propto\Gamma^{-1}[T_i,\mathbf{X}]D[T_i,\mathbf{X}]\cdot\mathbf{X}$, the corresponding optimized measurement sensitivity can be described as
\begin{align}
M[T_i,\mathbf{X}]=\textmd{max}_{\tilde{m}}(\partial_i\langle{X_{\tilde{m}}}\rangle)^2/(\delta {X_{\tilde{m}}})^2\\
=\textbf{D}[T_i,\mathbf{X}]^T\Gamma^{-1}[T_i,\mathbf{X}]\textbf{D}[T_i,\mathbf{X}],
\label{eq:A33}
\end{align}
where linear combinations $X_{\tilde{\textbf{m}}}=\tilde{\textbf{m}}\cdot\textbf{X}$, $\textbf{D}[T_i,\mathbf{X}]=(\partial_i\langle {X}_1\rangle,...,\partial_i\langle{X}_K\rangle)^T$ and the elements of the covariance matrix are $\Gamma_{k,l}[T_i,\mathbf{X}]=\langle{X}_k{X}_l\rangle-\langle {X}_k\rangle\langle{X}_l\rangle$.  The optimized sensitivity given by Eq.~(\ref{eq:A33}) is obtained by the measurement coefficients vector $\tilde{\textbf{m}}=\textbf{m}$.
The measurement sensitivity obeys the chain of inequalities $M[T_i,\mathbf{X}]\leq \mathcal{F}[T_i,X_\textbf{m}]\leq \mathcal{H}^{ii}$.
Here $\mathcal{F}[T_i,X_\textbf{m}]$ denotes the Fisher information (FI) of $T_i$ obtained from the measurement of ${X}_\textbf{m}$; $ \mathcal{H}^{ii}$ denotes the QFI of $T_i$ as shown in Eq.(24).

Photon counting after spatial-mode demultiplexing has been shown to be the measurement that allow one to approach the ultimate limit for the separation estimation\cite{lab6,lab26}. Supposing that we have access to $K$ orthonormal spatial modes $\{\upsilon_k(x)\}$ with associated field operators $a_k$ and that the photon number in each mode can be obtained from the photon counting operator $N_k=b_k^\dagger b_k$.
$b_k=g_{k+}a_++g_{k-}a_-$ with $g_{k\pm}=\int dx \upsilon^*_k(x)\psi_\pm (x)$. Then, the mean photon number in each mode is
\begin{align}
\langle N_k\rangle=N\eta(|f_{+,k}|^2+|f_{-,k}|^2)-\gamma N\eta(|f_{+,k}|^2-|f_{-,k}|^2),
\end{align}
where $f_{\pm,k}=\int dx \upsilon^*_k(x)\psi(x\pm d/2)$.

Next, we focus on the case of a Gaussian point spread function $\psi(x)=\sqrt{2/\pi\varpi^2}\exp(-x^2/\varpi^2)$.
For small average number of photons, demultiplexing Hermite-Gauss (HG) modes can help to approach the QCRB.
Hence, we also consider the orthonormal spatial modes
\begin{align}
\upsilon_k(x)=u_{k}(x)=\mathcal{N}_{k}H_n(\frac{\sqrt{2}x}{\varpi})e^{-x^2}
\end{align}
where $H_n(x)$ are the Hermite polynomials and the normalization constant $\mathcal{N}_{k}=[(\pi/2)\varpi^22^{k}k!]^{-1/2}$.

Let $X_k=N_k, \mathbf{X}=\mathbf{N}=(N_1,...,N_K)^T$, the measurement sensitivity of $T_i$ can be obtained by Eq.~(\ref{eq:A33})
\begin{align}
M[T_i,\mathbf{N}]=(2\eta\partial_iN)^2[\frac{\sum_{k=0}^K\beta_{k}^2(d)}{2N\eta}-\frac{A_+}{A_+A_--B^2}\mathcal{S}_1^2\nonumber\\
+\frac{2B}{A_+A_--B^2}\mathcal{S}_1\mathcal{S}_2-\frac{A_-}{A_+A_--B^2}\mathcal{S}_2^2],
\end{align}
where
\begin{align}
A_\pm=\frac{2}{1\pm\gamma^2}+2N\eta{\sum_{k=0}^K\beta_{k}^2(d)},\\
B=2N\eta\sum_{k=0}^K(-1)^{k}\beta_{k}^2(d),\\
\mathcal{S}_1=\sum_{k=0}^K(-1)^{k}\beta_{k}^2(d),\\
\mathcal{S}_2=\sum_{k=0}^K\beta_{k}^2(d)
\end{align}
Here, $\beta_{k}(d)=f_{\pm,k}=\frac{1}{\sqrt{k!}}\exp[\frac{d^2}{8\varpi^2}](\pm\frac{ d}{2\varpi})^k$.
When  the number of received photons is low $N\eta\ll1 $, the sensitivity can be simplified as
\begin{align}
M[T_i,\mathbf{N}]&\approx (2\eta\partial_iN)^2\frac{\sum_{k=0}^K\beta_{k}^2(d)}{2N\eta}\nonumber\\
&=\sum_{k=0}^K\frac{(\partial_iN_k)^2}{N_k}
=N_t\sum_{k=0}^K\frac{(\partial_ip_k)^2}{p_k}
\end{align}
where the total number of the thermal photons $N_t=\sum_{k=0}^KN_k$ and the probability $p_k=N_k/N_t$. The above equation shows that the FI is obtained, which means that the estimation strategy based on the optimized error transfer formula can saturate the Cram\'{e}r-Rao bound.
\begin{figure}
  \includegraphics[scale=0.72]{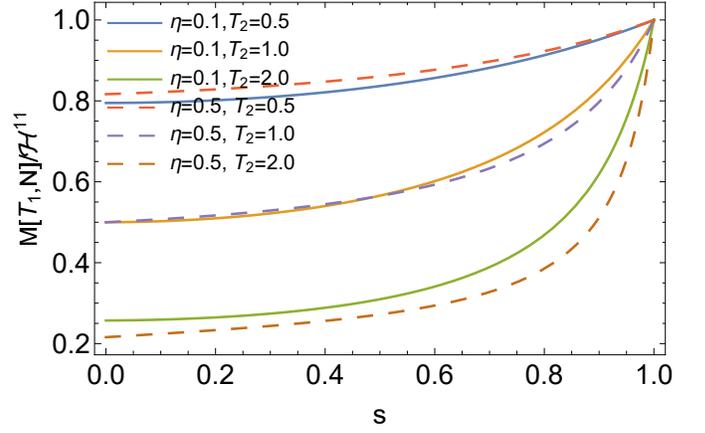}
  \caption{\label{fig.7}Diagram shows the ration of the measurement sensitivity and the QFI $M[T_1,\mathbf{N}]/\mathcal{H}^{11}$ versus the degree of diffraction $s$. Here, the values of the  selected dimensionless parameters are given by $\omega=1$, and $T_1=1$.}
\end{figure}

When the full HG basis is measured, i.e., $K\longrightarrow\infty$, we can obtain the sensitivity of the single parameter $T_i$
\begin{align}
&M[T_i,\mathbf{N}]=(2\eta\partial_iN)^2\nonumber\\
&[\frac{1}{2N\eta}-\frac{2s^2(1-\gamma^2)+2(1+\gamma^2)+2N\eta(1-\gamma^4)(s-1)^2}{4+8N\eta+4N^2\eta^2(1-s^2)(1-\gamma^4)}],
\end{align}
where $\partial_iN=\chi_i\omega/2T_i^2(1-\chi_i)^2$.
As shown in Fig.~\ref{fig.7}, the measurement sensitivity $M[T_1,\mathbf{N}]$ gradually approaches the QFI as the degree of the diffraction $s$ increases. In other words, as $s$ increases, the estimation strategy based on the optimized error transfer formula tends to be the optimal method by using the full HG basis.  At the maximum diffraction $s=1$, the estimation strategy can saturate the QCRB without being affected by the attenuation factor $\eta$.
\section{simultaneous estimation with a practical measurement operator}
In this section, we use a practical measurement to estimate the two parameters $T_1$ and $T_2$ simultaneously.
We consider a simple measurement operator $E=\sum_{k=0}^\infty N_k$, which is the direct sum of all the photon counting after the HG spatial-mode demultiplexing. It is independent of the estimation parameters $T_i$.

After a simple calculation, we can obtain that $E=\sum_{k=0}^\infty N_k=a_+^\dag a_++a_-^\dag a_-=(n_++n_-)|n_+,n_-\rangle\langle n_+,n_-|$. Conditioned on a detection event, the probability of detecting $n_+$ and $n_-$ photons in the modes of $a_+$ and $a_-$ is given by $P_{n_+,n_-}=\langle n_+,n_-|\rho|n_+,n_-\rangle$, which  can be further expressed as
\begin{align}
&P_{n_+,n_-}=\nonumber\\
&\frac{(1-\gamma^2)^{1+n_++n_-}N_+^{n_+}N_-^{n_-}
\ _2F_1[1+n_+,1+n_-,1,\frac{\gamma^2}{\lambda_+\lambda_-}]}{\lambda_+^{n_++1}\lambda_-^{n_-+1}},
\end{align}
where the parameters in the denominator are $\lambda_\pm=1+N_\pm-N_\pm\gamma^2$, and the Hypergeometric Function $\ _2F_1[1+n_+,1+n_-,1,\frac{\gamma^2}{\lambda_+\lambda_-}]=\sum_{m=0}^\infty\frac{[(1+n_+)(1+n_-)\frac{\gamma^2}{\lambda_+\lambda_-}]^m}{m!}$.

With this measurement probability, the FI can be calculated by
\begin{align}
\mathcal{F}_C^{ij}=\sum_{n_+,n_-=0}^\infty \frac{\partial_iP_{n_+,n_-}\partial_jP_{n_+,n_-}}{P_{n_+,n_-}}.
\end{align}

\begin{figure}
\includegraphics[scale=0.65]{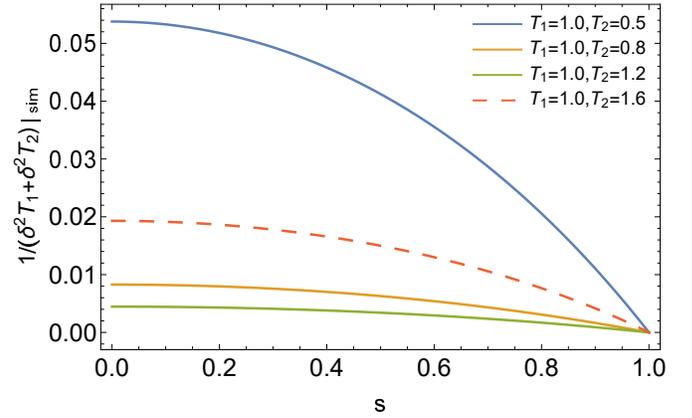}
   \caption{\label{fig.8}This plot depicts that the reciprocal of the simultaneous estimation uncertainty $\frac{1}{(\delta^2T_1+\delta^2T_2)|_{\textmd{sim}}}$ obtained by the measurement operator $E$  changes with the degree of diffraction $s$. Here, the values of the  selected dimensionless parameters are given by $\omega=1$, and $\eta=1/2$.}
\end{figure}
As shown in Fig.~\ref{fig.8}, the reciprocal of the simultaneous estimation uncertainty $\frac{1}{(\delta^2T_1+\delta^2T_2)|_{\textmd{sim}}}$ decreases with the degree of the diffraction $s$. When $s=1$, $\frac{1}{(\delta^2T_1+\delta^2T_2)|_{\textmd{sim}}}=0$. It shows that the diffraction will reduce the precision of the simultaneous estimation with the practical measurement operator $E$, which can not obtain any information at the maximum diffraction. This results  support the previous results using the saturated QCRB  as shown in section IV.B.

\section{conclusion}
We have investigated the effect of the diffraction on the quantum thermometry. When we know a priori that the temperatures of the two thermal sources are always equal, the diffraction will reduce the estimation precision but not by much: at low temperature, the diffraction has little effect on the estimation precision; at high temperature, the maximum diffraction yields half as much the QFI as no diffraction.
More QFI can be obtained with the priori information (the two temperatures are always equal) than that without the priori information (i.e., the two temperatures of the two thermal sources are independent). In particular, at the maximum diffraction, the prior information can help to obtain twice as much QFI as without the priori information. What's more, we carefully consider the two strategies: the simultaneous estimation and the individual estimation. The simultaneous estimation of two temperatures  is proved to satisfy the saturation condition of QCRB. Given the same resources, the simultaneous estimation performs better than the individual estimation in the case of small degree of diffraction. However, in the case of high degree of diffraction, the individual estimation performs better. In particular, at the maximum diffraction, the simultaneous estimation can not get any information, which is supported by a practical measurement, while the individual estimation can still get the information. In addition, we find that for the individual estimation, the practical and feasible estimation strategy based on the optimized error transfer formula can saturate the Cram\'{e}r-Rao bound when the number of received photons is low.  At the maximum diffraction, the practical and feasible estimation strategy by using the full HG basis can saturate the QCRB without being affected by the attenuation factor.

Our study illustrates the effect of diffraction on the temperature measurement precision and the advantages and disadvantages of different measurement strategies, which lays a foundation for constructing a remote precision thermometry.

\section*{Acknowledgements}
We acknowledge Qiongyi He for helpful discussion and constructive comments on the paper.
This research was supported by the National Natural Science Foundation of China under Grant No. 62001134, Guangxi Natural Science Foundation under Grant No. 2020GXNSFAA159047 and National Key R\&D Program of China under Grant No. 2018YFB1601402-2.

\section*{Appendix}

We now use Eq.~(\ref{eq:A09}) to propagate the density matrix $\rho_0$ of the sources to the density matrix $\rho$ in the image plane.

When we transform the coherent state $|\alpha_1,\alpha_2\rangle$ of the field operators $c_{1}$ and $c_2$ to the coherent state $|\alpha_+,\alpha_-\rangle$ of the field operators $a_\pm$, we can obtain the following mapping relation according to Eq.~(\ref{eq:A09}) and the auxiliary modes in the vacuum state
\begin{align}
\sqrt{\eta_+}c_\pm|\alpha_1,\alpha_2\rangle\langle\alpha_1,\alpha_2|\rightarrow a_\pm|\alpha_+,\alpha_-\rangle\langle\alpha_+,\alpha_-|\Rightarrow\nonumber\\
\sqrt{\eta_\pm/2}(\alpha_1\pm\alpha_2)|\alpha_1,\alpha_2\rangle\langle\alpha_1,\alpha_2|\rightarrow\alpha_\pm|\alpha_+,\alpha_-\rangle\langle\alpha_+,\alpha_-|.
\end{align}

Based on above equations, we obtain the mapping relations
\begin{align}
\alpha_{1}\rightarrow\alpha_{1+}=\alpha_+/\sqrt{2\eta_+}+\alpha_-/\sqrt{2\eta_-};\\
\alpha_{2}\rightarrow\alpha_{2-}=\alpha_+/\sqrt{2\eta_+}-\alpha_-/\sqrt{2\eta_-}.
\end{align}
Then, with the two equations above we further obtain
\begin{align}
&\int d^2\alpha_1d^2\alpha_2 P_{c_1,c_2}(\alpha_1,\alpha_2)|\alpha_1,\alpha_2\rangle\langle\alpha_1,\alpha_2|\rightarrow \nonumber\\
&\int d^2\alpha_{1+} d^2\alpha_{2-} P_{c_1,c_2}(\alpha_{1\pm},\alpha_{2\pm})|\alpha_+,\alpha_-\rangle\langle\alpha_+,\alpha_-|\\
&=\int d^2\alpha_+d^2\alpha_- P_{a_+,a_-}(\alpha_+,\alpha_-)|\alpha_+,\alpha_-\rangle\langle\alpha_+,\alpha_-|.
\end{align}
At this point, Eq.~(\ref{eq:A10}) has been derived.

\end{document}